\documentclass[12pt]{article}
\usepackage{amsmath}
\usepackage{amssymb}
\usepackage{amsfonts}
\usepackage{latexsym}
\usepackage{color}
\usepackage{graphicx}

\catcode `\@=11 \@addtoreset{equation}{section}

\catcode `\@=12

\newcommand{\be}{\begin{equation}}
\newcommand{\en}{\end{equation}}
\newcommand{\bea}{\begin{eqnarray}}
\newcommand{\ena}{\end{eqnarray}}
\newcommand{\beano}{\begin{eqnarray*}}
\newcommand{\enano}{\end{eqnarray*}}
\newcommand{\bee}{\begin{enumerate}}
\newcommand{\ene}{\end{enumerate}}

\newcommand{\Hil}{{\cal H}}

\newcommand{\F}{{\cal F}}
\newcommand{\Lc}{{\cal L}}
\newcommand{\Sc}{{\cal S}}
\newcommand{\D}{{\cal D}}

\newcommand{\E}{{\cal E}}

\newcommand{\1}{1 \!\! 1}

\begin{document}
\thispagestyle{empty}

\vspace*{1cm}

\begin{center}
{\Large \bf Some results on the rotated infinitely deep potential and its coherent states}   \vspace{2cm}\\

{\large F. Bagarello}\\
\vspace*{1cm}

\normalsize
Dipartimento di Ingegneria, Universit\`a di Palermo, I-90128  Palermo, Italy\\
and\\
INFN, Sezione di Napoli, Complesso Universitario di Monte S. Angelo,\\
Via Cintia Edificio 6, 80126 Napoli, Italy

\end{center}

\vspace*{0.5cm}

\begin{abstract}
\noindent The Swanson model is an exactly solvable model in quantum mechanics with a manifestly non self-adjoint Hamiltonian whose eigenvalues are all real. Its eigenvectors can be deduced easily, by means of suitable ladder operators. This is because the Swanson Hamiltonian is deeply connected with that of a standard quantum Harmonic oscillator, after a suitable rotation in configuration space is performed. In this paper we consider a rotated version of a different quantum system, the infinitely deep potential, and we consider some of the consequences of this rotation. In particular, we show that differences arise with respect to the Swanson model, mainly because of the technical need of working, here, with different Hilbert spaces, rather than staying in $\Lc^2(\mathbb{R})$. We also construct Gazeau-Klauder coherent states for the system, and  analyse their properties.
\end{abstract}

\vspace{2cm}

{\bf Keywords}:  Deformed quantum mechanical systems; Orthonormal bases; Gazeau-Klauder coherent states

\vfill

\newpage

\section{Motivation}\label{sect1}

The literature on quantum mechanics is extremely rich and it has produced  many results dealing with several aspects of the microscopic world,  both from the physical side and from a more mathematical point of view. In ordinary quantum mechanics it is usually assumed that the Hamiltonian $H$ of a given physical system $\Sc$, driving its dynamics, is self-adjoint: $H=H^\dagger$. The same is assumed for the other observables of $\Sc$, e.g. for the position or the momentum operators of the particles in $\Sc$. This choice is adopted also because, in this way, the mean values of all these operators, and their eigenvalues in particular, are real and, therefore, directly connected with some experiment which can be performed on $\Sc$. Also, the dynamics is unitary, and the probabilistic interpretation of the wave function is preserved over  time. This is the standard point of view widely considered in many textbooks, \cite{mer}-\cite{cohen}.

In recent years people started to be interested in the possibility of using observables which are not self-adjoint, since some of them can still have real eigenvalues and produce an unitary dynamics. We refer to \cite{bagbook}-\cite{mosta} for many results on what is usually called $PT$- or {\em pseudo-hermitian} quantum mechanics\footnote{A plethora of other different names have also been used along the years for this extended quantum mechanics, by different authors, see also \cite{milosh}.}. One of the relevant aspects of these approaches is that they provide examples of manifestly non-selfadjoint Hamiltonians whose eigenvectors are real. One such an Hamiltonian was proposed by Swanson in \cite{swan}, and then considered by other authors, \cite{dapro}-\cite{baginbagbook}. The Hamiltonian for the model can be written as
\be
H_\theta=\frac{1}{2\cos 2\theta}\left[\left(e^{-i\theta}p\right)^2+\left(e^{i\theta}x\right)^2\right]. \label{Htheta}
\en  
Thus, $H_\theta$ is obtained from the standard harmonic oscillator $H_{\theta=0}$ (with spectrum $E_n=n+\frac{1}{2}$) by a complex canonical transformation
$
x\rightarrow e^{i\theta} x$ and $p\rightarrow e^{-i\theta}p,
$ the second being a consequence of the first since $p$ is proportional to the first space derivative, $\frac{d}{dx}$. As discussed in \cite{bagpl2010}, $H_\theta$ can be rewritten in terms of the so-called pseudo-bosonic operator $A_\theta$ and $B_\theta$,
$$\left\{
\begin{array}{ll}A_\theta=\frac{1}{\sqrt{2}}\left(e^{i\theta}x+e^{-i\theta}\,\frac{d}{dx}\right),\\
	B_\theta=\frac{1}{\sqrt{2}}\left(e^{i\theta}x-e^{-i\theta}\,\frac{d}{dx}\right),\end{array} \right. $$
	which satisfy the commutation rule $ [A_\theta,B_\theta]=\1$, \cite{bagpl2010,baginbagbook}. In terms of these operators we find that $H_\theta=\frac{1}{\cos 2\theta}\left[ B_\theta A_\theta+\frac{1}{2}\,\1\right]$. Since $ B_\theta A_\theta$ is a number-like operator, the eigenvalues of $H_\theta$ are essentially those of the harmonic oscillator, and the eigenstates can be written in terms of rotated Hermite polynomials times the usual gaussian function $e^{-x^2/2}$, again rotated of the angle $\theta$, $e^{-e^{2i\theta}x^2/2}$. The set of these eigenstates is complete in $\Lc^2(\mathbb{R})$, but it is not an orthonormal (o.n.) basis. Actually, in \cite{bagpl2010} it was proved that this set is not even a basis. However, a second set of functions can be introduced, biorthogonal to the eigenstates of $H_\theta$, which are eigenstates of $H_\theta^\dagger$. Again, this set is complete in $\Lc^2(\mathbb{R})$, but it is not a basis.  A special class of coherent states can be constructed, having most of the properties of standard coherent states. These vectors have been called {\em bi-coherent} states, and are eigenstates of $A_\theta$ and $B_\theta^\dagger$, respectively. This particular aspect of Swanson model has been considered in \cite{bgs,bgs2}.
	
	\vspace{2mm} 
	
	{\bf Remark:--} It might be useful to notice that some authors have considered what happens to $H_0$ in (\ref{Htheta}) if, rather than rotating the variables one considers complex translations. This has been done, for instance, in \cite{bag2013,fakhri,petr}.
	
	\vspace{2mm} 
	
	This extremely synthetic summary of \cite{bagpl2010} shows how the mathematical and the physical consequences of a rotation $x\rightarrow e^{i\theta} x$ are quite rich and interesting, and motivates our analysis here, where again we consider what happens when we {\em rotate} a well known and well studied quantum system, an infinitely deep symmetric wall potential (IDSWP). Contrarily to what happens in the Swanson model, in our rotated version of the IDSWP we will use a continuous family of Hilbert spaces, labeled by the rotation angle, and we will find o.n. bases for each of these spaces. Isospectral self-adjoint Hamiltonians will be produced,  and some of their properties and of the properties of their eigenstates will be analyzed. Also, we will define operators between different Hilbert spaces which are unitary or unbounded, depending on the spaces where they act. In Section \ref{sectcs} we introduce  coherent states of the Gazeau-Klauder type for the model, and we study their properties. Conclusions and perspectives are given in Section \ref{sectconclu}

\section{Rotating the wall}

In this section we first briefly review few facts for the IDSWP, and then we discuss the effect of the rotation on the system. In the following we will mainly work in the Hilbert space $\Hil_0=\Lc^2\left(-\frac{L}{2},\frac{L}{2}\right)$, $L>0$ fixed, endowed with the scalar product $\langle f,g\rangle=\int_{-L/2}^{L/2}\overline{f(x)}\,g(x)\,dx$, $f(x), g(x)\in\Hil_0$.

\subsection{The unrotated wall}\label{sectunrowall}

The Hamiltonian $H_0$ we deal with is the following:
\be
(H_0 f)(x)=\left(-\frac{d^2}{dx^2}+V(x)\right)f(x),
\label{21}
\en
where $V(x)=0$ if $x\in\left[-\frac{L}{2},\frac{L}{2}\right]$, and $V(x)=\infty$ outside this interval. Here
 $f(x)\in D(H_0)$, which is defined as follows:
\be
D(H_0)=\left\{f(x)\in\Hil_0, \mbox{ twice differentiable: } f''(x)\in\Hil_0 \mbox{ and } f\left(\pm\frac{L}{2}\right)=0\right\}.
\label{22}
\en
This choice of the boundary conditions are motivated from physics, and in particular from the impossibility of the particle to enter the zone of the real line where the potential energy is infinity. From a mathematical point of view, it is easy to check that, with this choice, $H_0=H_0^\dagger$. The eigenvalues and the eigenvectors of $H_0$ can be found in many textbooks, see \cite{mess}, for instance. We have
\be
\Phi_j(x)=\sqrt{\frac{2}{L}}\left\{
\begin{array}{ll}\sin\left(\frac{2k\pi x}{L}\right),\quad \qquad \mbox{ if } j=2k\\
	\cos\left(\frac{(2k-1)\pi x}{L}\right), \quad \mbox{ if } j=2k-1,\end{array} \right.
\label{23}\en
while the eigenvalues are $E_j=\left(\frac{j\pi}{L}\right)^2$, for all $j\geq1$: $H_0\Phi_j(x)=E_j\Phi_j(x)$. So we have odd and even eigenstates for $H_0$. They are all mutually orthogonal. In fact, we have
\be
\langle \Phi_j,\Phi_k\rangle=\delta_{j,k}.
\label{24}\en
We call $\F_{\Phi}=\{\Phi_j(x), \,j\geq1\}$ the set of all these functions. $\F_{\Phi}$ is an o.n. basis for $\Hil_0$, as one can deduce from Proposition 3.24 of \cite{hall}.

\subsection{The rotation}

Let us now introduce the following rotation operator $T_\varphi$, $\varphi\in\mathbb{R}$ fixed, on each function $\Phi_j(x)$:
\be
\Phi_j^{(\varphi)}(x):=T_\varphi\Phi_j(x)=\Phi_j(e^{i\varphi}x).
\label{25}\en
Using simple trigonometric formulas, and the connection between them and the hyperbolic sinus and cosinus, it is easy to compute the real and the imaginary parts of the various $\Phi_j^{(\varphi)}(x)$. For instance, we have
$$
\Re\left\{\sin\left(\frac{2k\pi e^{i\varphi}x}{L}\right)\right\}= \sin\left(\frac{2k\pi x}{L}\,\cos(\varphi)\right)\cosh\left(\frac{2k\pi x}{L}\,\sin(\varphi)\right),
$$
and
$$
\Im\left\{\sin\left(\frac{2k\pi e^{i\varphi}x}{L}\right)\right\}= \cos\left(\frac{2k\pi x}{L}\,\cos(\varphi)\right)\sinh\left(\frac{2k\pi x}{L}\,\sin(\varphi)\right),
$$
and similar expressions can be deduced for the real and imaginary parts of $\cos\left(\frac{(2k-1)\pi e^{i\varphi}x}{L}\right)$. In other words, the result of the rotation of the functions in (\ref{25}) can be rewritten as a simple (complex) combination of real functions. We can easily extend $T_\varphi$ to products of trigonometric and hyperbolic functions, since similar formulas can be used. In other words, we can rotate twice the functions $\Phi_j(x)$, also of different angles. In particular, an useful identity is the following
\be
T_\varphi\Phi_j(x)=T_\alpha T_\beta \Phi_j(x),
\label{26}\en
for all real $\alpha$ and $\beta$ such that $\alpha+\beta=\varphi$. The geometric meaning of this equality is clear: two successive rotations of $\beta$ and $\alpha$ act as a single rotation of $\alpha+\beta$. An analytic proof can be deduced using well known identities for trigonometric and hyperbolic functions, as those already given.

It is interesting to compute the norm of the various $\Phi_j(x)$'s in $\Hil_0$,  to see that they are  finite for all $j$. This implies that each $\Phi_j(x)$ is in the domain of the operator $T_\varphi$, $\forall\, \varphi$, which however, as we will show in a moment, cannot be all of $\Hil_0$. We have
$$
\|\Phi_j^{(\varphi)}\|^2=\left\{\begin{array}{ll} \frac{1}{2k\pi}\left(\csc\varphi\,\sinh(2k\pi\sin\varphi)-\sec\varphi\,\sin(2k\pi\cos\varphi)\right),
 \qquad\qquad\qquad\quad \mbox{ if } j=2k\\
\frac{1}{2(2k-1)\pi}\left(\csc\varphi\,\sinh((2k-1)\pi\sin\varphi)+\sec\varphi\,\sin((2k-1)\pi\cos\varphi)\right), \, \mbox{ if } j=2k-1.\end{array} \right.
$$
It is interesting to notice that in both these cases, $j$ odd or $j$ even, $\|\Phi_j^{(\varphi)}\|$ converges to 1 when $\varphi\rightarrow0$, as it should. It is also interesting to observe that, if $\varphi\neq\ k\pi$, $k\in\mathbb{Z}$, formula above shows that $T_\varphi$ is unbounded. In fact, $\|\Phi_{j}^{(\varphi)}\|$ diverges exponentially with $j$, both for even and for odd $j$. For this reason, we introduce the (maximal) domain of $T_\varphi$, $D(T_\varphi)$, as follows:
$$
D(T_\varphi)=\left\{f(x)\in\Hil_0: \sum_j\langle\Phi_j,f\rangle\Phi_j^{(\varphi)}(x) \in\Hil_0\right\},
$$
and $\forall f(x)\in D(T_\varphi)$, $f(x)=\sum_j\langle\Phi_j,f\rangle\Phi_j(x)$, we have $T_\varphi f(x)=\sum_j\langle\Phi_j,f\rangle\Phi_j^{(\varphi)}(x)$. Of course, $\Phi_k(x)$ belongs to this set, for all $k$, as well as their finite linear combinations. Hence $D(T_\varphi)$ is dense in $\Hil_0$, and $T_\varphi$ is densely defined.

Using the identity in (\ref{26}), the following results can be proved: 
\be
\Phi_j^{(\varphi)}(x)\in D(T_{-\varphi})\cap D(T_{-\varphi}^2), \quad\mbox{ and }\quad T_{-\varphi}\Phi_j^{(\varphi)}(x)=\Phi_j(x), \quad T_{-\varphi}^2\Phi_j^{(\varphi)}(x)=\Phi_j^{(-\varphi)}(x),
\label{27}\en
where $D(T_{-\varphi}^2)$ is defined in analogy with $D(T_{\varphi})$ above:
$$
D(T_{-\varphi}^2)=\left\{g(x)\in\Hil_0: \sum_j\langle\Phi_j,g\rangle\Phi_j^{(-2\varphi)}(x) \in\Hil_0\right\}.
$$
In fact, 
$$
 T_{-\varphi}\Phi_j^{(\varphi)}(x)= T_{-\varphi}T_{\varphi}\Phi_j(x)=\Phi_j(x), \quad  T_{-\varphi}^2\Phi_j^{(\varphi)}(x)= T_{-\varphi}^2T_{\varphi}\Phi_j(x)=T_{-\varphi}\Phi_j(x)=\Phi_j^{(-\varphi)}(x).
$$
It is easy to understand that $T_\varphi$ is not unitary as an operator from $\Hil_0$ into itself, for general values of $\varphi$. This is because the norm of $\Phi_j^{(\varphi)}(x)$, $\|\Phi_j^{(\varphi)}\|$, is different 
from 1. Hence,  for these values of $\varphi$, $T_\varphi$ does not preserve the norm of $\Hil_0$ and, consequently, cannot be unitary as an operator from $\Hil_0$ into itself. However, how we will show later, $T_\varphi$ is in fact unitary, but as an operator between different Hilbert spaces. We will come back on this aspect of $T_\varphi$ in Section \ref{sectOHS}. Furthermore,  it is also possible to check that $T_\varphi$ cannot be self-adjoint in $\Hil_0$. In fact, let us suppose that this is true. Then, since both $\Phi_k^{(\varphi)}$ and $\Phi_j^{(-\varphi)}$ belong to $\Hil_0$, we should have
$$
\langle \Phi_k^{(\varphi)},\Phi_j^{(-\varphi)}\rangle=\langle T_{\varphi}\Phi_k,T_{-\varphi}\Phi_j\rangle=\langle\Phi_k, T_{\varphi}T_{-\varphi}\Phi_j\rangle=\langle\Phi_k, \Phi_j\rangle=\delta_{k,j},
$$
which would imply that the families $\F_{\Phi}^{(\varphi)}=\{\Phi_j^{(\varphi)}(x), \,j\geq1\}$ and $\F_{\Phi}^{(-\varphi)}=\{\Phi_j^{(-\varphi)}(x), \,j\geq1\}$ are biorthonormal. Indeed, this is false, as a direct computation easily shows: in fact, we can check that, for instance, 
$$
\langle \Phi_2^{(\varphi)},\Phi_4^{(-\varphi)}\rangle=\frac{4e^{i\varphi}}{3\pi}\sin^3(e^{-i\varphi}\pi),
$$
which is different from zero, in general. This formula also shows that, if $\varphi=0$ (i.e., for the unrotated wall), the scalar product is zero, as it should. The conclusion is therefore that, as stated, $T_\varphi$ is different from $T_\varphi^\dagger$.

Incidentally, this analysis suggests how to construct, at least in principle, a family of vectors  $\F_{\eta}^{(\varphi)}=\{\eta_j^{(\varphi)}(x), \,j\geq1\}$ which is biorthogonal to  $\F_{\Phi}^{(\varphi)}$: it is sufficient to define $\eta_j^{(\varphi)}(x)=T_\varphi^\dagger\,\Phi_j(x)$, whenever this is a well-defined operation. However, due to the particular form of $T_\varphi$, the computation of its adjoint does not provide a really explicit expression, as we will show in a moment.

A standard exercise shows that $$D(T_\varphi^\dagger)=\{g(x)\in\Hil_0: \sum_j|\langle \Phi_j^{(\varphi)},g\rangle|^2<\infty\},$$ and $$T_\varphi^\dagger \,g(x)=\sum_j\langle  \Phi_j^{(\varphi)},g\rangle\Phi_j(x),$$ for all $g(x)\in D(T_\varphi^\dagger)$. We see immediately that it is not evident if, for generic $\varphi$,  $T_\varphi^\dagger$ is densely defined or not. 
However,  if the set $\F_{\Phi}^{(\varphi)}$ is a basis for $\Hil_0$ (but this is not granted), then $D(T_\varphi^\dagger)$ is dense in $\Hil_0$. Suppose, in fact, that this is the case, i.e. that $\F_{\Phi}^{(\varphi)}$ is a basis for $\Hil_0$. Then an unique biorthogonal set exists $\F_{\tilde\eta}^{(\varphi)}=\{\tilde\eta_j^{(\varphi)}(x), \,j\geq1\}$, which is also a basis for $\Hil_0$, \cite{chri}. In this case, it is clear that $D(T_\varphi^\dagger)$ is dense in $\Hil_0$, since it contains all the finite linear combinations of the $\tilde\eta_j^{(\varphi)}(x)$'s, which form a dense set in $\Hil_0$. Summarizing: if $\F_{\Phi}^{(\varphi)}$ is a basis for $\Hil_0$, then $T_\varphi^\dagger$ is densely defined.

\vspace{2mm}

{\bf Remarks:--} (1) Due to the fact that $T_\varphi$ is unbounded, the set $\F_{\Phi}^{(\varphi)}$ could be a basis, but it is surely {\bf not } a Riesz basis, since it is the image of an o.n. basis via an unbounded operator.

(2) The biorthogonal set $\F_{\tilde\eta}^{(\varphi)}$, if it exists, is surely different from $\F_{\Phi}^{(-\varphi)}$: $\tilde\eta_j^{(\varphi)}(x)\neq \Phi_j^{(-\varphi)}$. In fact, we have already seen that $\F_{\Phi}^{(-\varphi)}$ is not biorthogonal to $\F_{\Phi}^{(\varphi)}$, since $T_\varphi\neq T_\varphi^\dagger$. This would be enough to conclude. But we have more than this: since $\|\Phi_j^{(-\varphi)}\|\|\Phi_j^{(\varphi)}\|\rightarrow\infty$ for $j$ divergent, the series $\sum_j\langle\Phi_j^{(-\varphi)},f\rangle\Phi_j^{(\varphi)}(x)$ cannot be convergent for all $f(x)\in\Hil_0$, \cite{dav}. This is again against the possibility that $\F_{\Phi}^{(\varphi)}$ and $\F_{\Phi}^{(-\varphi)}$ are biorthogonal bases.

\vspace{2mm}

A detailed analysis of $T_\varphi^\dagger$ will be considered in a future paper, since this is not essential here. In fact, we have already seen that some useful results can still be deduced also without an explicit knowledge of $T_\varphi^\dagger$. In fact, we expect that the analysis of $T_\varphi$, $T_\varphi^{-1}$ and $T_\varphi^\dagger$ can give rise to interesting features, mainly on the mathematical side.

\subsection{Other Hilbert spaces}\label{sectOHS}

As we have already said before, it is useful to work also with different Hilbert spaces other than $\Hil_0$. The reason for this is that, for instance, the operator $T_\varphi$ will acquire particularly nice mathematical properties when considered as an operator between suitable different Hilbert spaces.

We start defining the set $\Lc_\Phi=l.s.\{\Phi_j(x)\}$, the finite linear span of the functions in (\ref{23}). Of course, $\Lc_\Phi$ is dense in $\Hil_0$. Any function $f(x)\in\Lc_\Phi$ is a  linear expansion of the form
\be
f(x)=\sum_kc_k(f)\Phi_k(x), \qquad c_k(f)=\langle\Phi_k,f\rangle,
\label{27bis}\en
where the sum is finite and where we have written explicitly the dependence on $f$ of the coefficients of the expansion $c_k(f)$. To any such function we can associate another function, still in $\Hil_0$, in the following way:
\be
f^{(\varphi)}(x):=T_\varphi f(x)=\sum_kc_k(f)\Phi_k^{(\varphi)}(x),
\label{28}\en
where the coefficients $c_k(f)$ are those in (\ref{27bis}). We call  $$\Lc_\Phi^{(\varphi)}=\{h(x)\in\Hil_0: \mbox{ for some } h_0(x)\in\Lc_\Phi, \quad h(x)=T_\varphi h_0(x)\}.$$ In other words, $\Lc_\Phi^{(\varphi)}$ is just the image of $\Lc_\Phi$ via $T_\varphi$. In fact, since any $h_0(x)\in\Lc_\Phi$ can be written as $h_0(x)=\sum_kc_k(h_0)\Phi_k(x)$, $h(x)$ must necessarily have the following form: \be h(x)=T_\varphi h_0(x)=T_\varphi \sum_kc_k(h_0)\Phi_k(x)=\sum_kc_k(h_0)\Phi_k^{(\varphi)}(x).\label{29}\en
This means that $\Lc_\Phi^{(\varphi)}$ is just the finite linear span of the $\Phi_k^{(\varphi)}(x)$:  $\Lc_\Phi^{(\varphi)}=l.s.\{\Phi_j^{(\varphi)}(x)\}$, which is still another way to look at $\Lc_\Phi^{(\varphi)}$.

 On this space we can introduce a new scalar product as follows:
\be
\langle f,g\rangle_\varphi=\langle T_{-\varphi}f,T_{-\varphi}g\rangle,
\label{210}\en
for all $f(x),g(x)\in\Lc_\Phi^{(\varphi)}$. Notice that this is a well defined scalar product since, first of all, by the definition of $\Lc_\Phi^{(\varphi)}$, both $T_{-\varphi}f$ and $T_{-\varphi}g$ belong to $\Hil_0$. The properties of scalar products are also easily verified. In particular, if $f(x)\in\Lc_\Phi^{(\varphi)}$ is such that $\langle f,f\rangle_\varphi=0$, then $T_{-\varphi}f(x)=0$ almost everywhere in $\left[-\frac{L}{2},\frac{L}{2}\right]$, which implies, because of the equation (\ref{26})\footnote{Since $f(x)\in\Lc_\Phi^{(\varphi)}$, $f(x)$ is a finite linear combination of functions $\Phi_k^{(\varphi)}(x)=T_\varphi\Phi_k(x)$. Hence (\ref{26}) can be used safely.}, that $f(x)=0$. Hence we can define a norm on $\Lc_\Phi^{(\varphi)}$ as usual, $\|f\|_\varphi^2=\langle f,f\rangle_\varphi$, and complete $\Lc_\Phi^{(\varphi)}$ with respect to this norm. We get an Hilbert space, $\Hil_\varphi$, which is different, in general, from $\Hil_0$. Of course, by construction, $\Lc_\Phi^{(\varphi)}$ is dense in $\Hil_\varphi$ in the norm $\|.\|_\varphi$. This is in complete analogy with $\Lc_\Phi$, which is dense in $\Hil_0$ in the norm $\|.\|$. Furthermore, it is easy to check that the set $\F_{\Phi}^{(\varphi)}$ is an o.n. basis for $\Hil_\varphi$. In fact, due to (\ref{210}), 
\be
\langle \Phi_k^{(\varphi)},\Phi_j^{(\varphi)}\rangle_\varphi=\langle T_{-\varphi}\Phi_k^{(\varphi)},T_{-\varphi}\Phi_j^{(\varphi)}\rangle=\langle \Phi_k,\Phi_j\rangle=\delta_{k,j}.
\label{211}\en
Completeness of $\F_\Phi^{(\varphi)}$ then follows from the definition of $\Hil_\varphi$, since, as we have seen, $\Lc_\Phi^{(\varphi)}$ is  the finite linear span of the $\Phi_k^{(\varphi)}(x)$. 

\vspace{2mm}

{\bf Remark:--} Since $\Lc_\Phi^{(\varphi)}$ is an o.n. basis for $\Hil_\varphi$, any function $f_\varphi(x)$ in $\Lc_\Phi^{(\varphi)}$  should admit the following expansion: $f_\varphi(x)=\sum_kc_k'(f_\varphi)\Phi_k^{(\varphi)}(x)$, with $c_k'(f_\varphi)=\langle\Phi_k^{(\varphi)},f_\varphi\rangle_\varphi$. But, since $T_{-\varphi}\Phi_k^{(\varphi)}(x)=\Phi_k(x)$, using (\ref{210}) we find $c_k'(f_\varphi)=\langle\Phi_k,T_{-\varphi}f_\varphi\rangle=\langle\Phi_k,f\rangle=c_k(f)$, see (\ref{27bis}), where we have introduced the function $f(x)=T_{-\varphi}f_\varphi(x)$, which belongs to $\Lc_\Phi$. This means that the two functions $f(x)$ and $f_\varphi(x)$ have the same expansion coefficients or, stated defferently, that these coefficients are {\em covariant}, i.e., they do not depend on the choice of $\varphi$ which fix the Hilbert space $\Hil_\varphi$.

\vspace{2mm}

A consequence of this analysis is the following: $T_\varphi$ maps the set $\F_\Phi$, which is an o.n. basis for $\Hil_0$, in  $\F_\Phi^{(\varphi)}$, which is also an o.n. basis, but for $\Hil_\varphi$. Hence, $T_\varphi$ is an unitary operator from $\Hil_0$ to $\Hil_\varphi$ and, therefore, its related norm is one:
\be
\|T_\varphi\|_{B(\Hil_0,\Hil_\varphi)}=1.
\label{212}\en
This result is true for each fixed $\varphi$. In particular, it is true if we take $\varphi=\theta$, for some positive $\theta$, and when we then take $\varphi=-\theta$: $T_\theta$ is an unitary operator from $\Hil_0$ to $\Hil_\theta$ with $\|T_\theta\|_{B(\Hil_0,\Hil_\theta)}=1$, while $T_{-\theta}$ is an unitary operator from $\Hil_0$ to $\Hil_{-\theta}$ with $\|T_{-\theta}\|_{B(\Hil_0,\Hil_{-\theta})}=1$. Moreover, it is easy to see that $T_\theta^2$ maps $\F_\Phi^{(-\theta)}$ into $\F_\Phi^{(\theta)}$ so that, for the same reasons, $T_\theta^2$
is an unitary operator from $\Hil_{-\theta}$ to $\Hil_{\theta}$ with $\|T_{\theta}^2\|_{B(\Hil_{-\theta},\Hil_{\theta})}=1$. Similarly,  $T_{-\theta}^2$ 
is an unitary operator from $\Hil_{\theta}$ to $\Hil_{-\theta}$ with $\|T_{-\theta}^2\|_{B(\Hil_{\theta},\Hil_{-\theta})}=1$.

\vspace{2mm}

{\bf Remark:--} Of course, the fact that $T_\varphi$ is unitary from $\Hil_0$ to $\Hil_\varphi$ does not help us in getting the explicit form of $T_\varphi^\dagger$, the adjoint of $T_\varphi$ in $\Hil_0$ since, at is is well known, the concept of adjoint is deeply connected to the scalar product we are working with. In other words, it is not true that $T_\varphi^\dagger=T_\varphi^{-1}$, while it is true that $T_\varphi^{\dagger_\varphi}=T_\varphi^{-1}$, where $T_\varphi^{\dagger_\varphi}$ is the adjoint of $T_\varphi$ in $\Hil_\varphi$, see Section \ref{sectTH}.

\subsection{The Hamiltonians}\label{sectTH}

Let us now introduce  an operator $H_\varphi$ as follows:
\be
H_\varphi=T_\varphi H_0 T_{-\varphi}.
\label{213}\en
The rigorous definition of its domain  will be given below. However, it is clear that this operator maps $\Lc_\Phi^{(\varphi)}$ into itself. Therefore $H_\varphi$ is densely defined in $\Hil_\varphi$.

 In fact, taken $f(x)=\sum_kc_k(f)\Phi_k^{(\varphi)}(x)=T_\varphi(\sum_kc_k(f)\Phi_k(x))\in \Lc_\Phi^{(\varphi)}$ and using equation (\ref{27}) and the results of Section \ref{sectunrowall}, we find the following:
$$
H_\varphi f(x)= H_\varphi\left(\sum_kc_k(f)\Phi_k^{(\varphi)}(x)\right)=T_\varphi H_0 T_{-\varphi}T_\varphi\left(\sum_kc_k(f)\Phi_k(x)\right)=$$
$$=\sum_kc_k(f)T_\varphi H_0\Phi_k(x)=\sum_kc_k(f)E_kT_\varphi\Phi_k(x)=\sum_kc_k(f)E_k\Phi_k^{(\varphi)}(x),
$$
which is again an element of $\Lc_\Phi^{(\varphi)}$. We recall that, in this derivation, all the sums are finite. Of course, if we fix $c_k(f)=\delta_{k,j}$, the above formula also shows in particular that $\Phi_j^{(\varphi)}(x)$ is an eigenstate of $H_\varphi$ with eigenvalue $E_j$: $H_\varphi\Phi_j^{(\varphi)}(x)=E_j\Phi_j^{(\varphi)}(x)$. Hence  $H_0$ and $H_\varphi$ have the same eigenvalues. This is not surprising, due to the relation between them, see (\ref{213}). In particular we see that, independently of the possible self-adjointness of $H_\varphi$ (in any of the spaces considered here), its eigenvalues are all real. We will return on this aspect later on.


It is easy to check that the map in (\ref{213}) changes the kinetic part of $H_0$,  $K_0=-\frac{d^2}{dx^2}$, into $K_\varphi=-e^{-2i\varphi}\frac{d^2}{dx^2}$, which is in agreement with what happens for the Swanson model, as discussed in Section \ref{sect1}. Notice that the full Hamiltonian, $H_\varphi$, changes as follows: $H_\varphi=K_\varphi+V(e^{i\varphi}x)$.

The fact that $H_\varphi$ admits an o.n. basis of eigenvectors may suggest that $H_\varphi$ is self-adjoint in $\Hil_\varphi$. In fact, it is possible to prove the following equality, which is a first step in proving this aspect of $H_\varphi$:
\be
\langle H_\varphi f,g\rangle_\varphi=\langle  f,H_\varphi g\rangle_\varphi,
\label{215}\en
for all $f(x), g(x)\in \Lc_\Phi^{(\varphi)}$. Indeed we have, using (\ref{210}) together with the fact that both $H_\varphi f(x)$ and $g(x)$ belong to $\Lc_\Phi^{(\varphi)}$,
$$
\langle H_\varphi f,g\rangle_\varphi=\langle T_{-\varphi}H_\varphi f,T_{-\varphi}g\rangle=\langle H_0T_{-\varphi} f,T_{-\varphi}g\rangle =\langle T_{-\varphi} f,H_0T_{-\varphi}g\rangle=\langle T_{-\varphi} f,T_{-\varphi}H_\varphi g\rangle,
$$
and (\ref{215}) follows. As already claimed, we can prove more than this: $H_\varphi$ is self-adjoint in $\Hil_\varphi$. This is a consequence of the fact that $H_\varphi$ has real eigenvalues and that its eigenvectors form an o.n. basis in $\Hil_\varphi$. In particular we can check that 
$$
D(H_\varphi)=\left\{f(x)=\sum_k c_k\Phi_k^{(\varphi)}(x):\, \sum_k |c_kE_k|^2<\infty\right\},
$$
coincides with $D(H_\varphi^{\dagger_\varphi})$: $\D_\varphi:=D(H_\varphi)=D(H_\varphi^{\dagger_\varphi})$, and that $H_\varphi f(x)=H_\varphi^{\dagger_\varphi}f(x)$ for all $f(x)\in D_\varphi$. Here we have used $\dagger_\varphi$ to indicate the adjoint in $\Hil_\varphi$: $\langle Af,g\rangle_\varphi=\langle f,A^{\dagger_\varphi}g\rangle_\varphi$, for all $f, g$ and $A$ for which this formula makes sense.

\vspace{2mm}

{\bf Remark:--} Notice that the above expression of $D(H_\varphi)$ is in agreement with what stated before, i.e. that $\Lc_\Phi^{(\varphi)}\subseteq D(H_\varphi)$. In fact, for all $f(x)\in\Lc_\Phi^{(\varphi)}$, the series $ \sum_k |c_kE_k|^2$ reduces to a finite sum, which is obviously bounded.

\vspace{2mm}

If we restrict to $\Hil_0$, rather than to $\Hil_\varphi$, we get the following result, which relates
 $K_\varphi$ with $K_{-\varphi}$:
\be
\langle K_{-\varphi} f,g\rangle=\langle  f,K_{\varphi}g\rangle,
\label{216}\en
for all $f(x), g(x)\in D(H_0)$. The proof is based on a double integration by part and on the use of the boundary conditions for functions belonging to $D(H_0)$, see (\ref{22}). 

\vspace{2mm}

{\bf Remark:--} It is interesting to observe that in (\ref{216})  the same kinetic terms appear in both sides of the equality if $\varphi=0$, and therefore the same Hamiltonians, since $V(e^{i\varphi}x)=V(e^{-i\varphi}x)=V(x)$, if $\varphi=0$. This is in agreement with the fact that, as we have discussed in Section \ref{sectunrowall}, $H_0$ is self-adjoint. It is also in agreement with a simple-minded view to the operator $K_\varphi$. In fact, (\ref{216}) suggests that $K_{-\varphi}^\dagger=K_\varphi$, which is exactly what one deduces simply rewriting $K_\varphi=2e^{-2i\varphi}p^2$ and recalling that $p=p^\dagger$: $K_\varphi^\dagger=2e^{2i\varphi}p^2=K_{-\varphi}$.

%
%
%
%

\section{Gazeau-Klauder-like coherent states}\label{sectcs}

As it is often true for many quantum systems, it is possible to {\em attach} coherent states also to our IDSWP, both in its original and in its rotated form. It is convenient to work directly with this last version, and therefore in $\Hil_\varphi$, since we can recover the unrotated version of coherent states simply by sending $\varphi$ to zero. 

It is well known that coherent states exist of many different kind, with similar but not identical properties, and which can be introduced using often slightly different procedures. We refer to \cite{gazeaubook}-\cite{cpcs} for some volumes on coherent states. The approach we will adopt here is based on the paper \cite{gk}, where the reader can find more details, included the analysis of Hamiltonians with continuous spectra. Our starting point is the following shifted version of $H_\varphi$: $h_\varphi=H_\varphi-E_1\1$, where $E_1=\left(\frac{\pi}{L}\right)^2$, see Section  \ref{sectunrowall}. If we call, for later convenience,
\be
\psi_k^{(\varphi)}(x)=\Phi_{k+1}^{(\varphi)}(x), \qquad \E_k=E_{k+1}-E_1=\left(\frac{\pi}{L}\right)^2\,k(k+2),
\label{31}\en
we have 
\be
h_\varphi\psi_k^{(\varphi)}(x)=\E_k\,\psi_k^{(\varphi)}(x),
\label{32}\en
$k=0,1,2,3,\ldots$, with $0=\E_0<\E_1<\E_2<\cdots$ and $\langle\psi_k^{(\varphi)},\psi_n^{(\varphi)}\rangle_\varphi=\delta_{k,n}$. Hence we are exactly in the settings of \cite{gk}, and we can define a family of coherent states as follows:
\be
\Psi^{(\varphi)}(J,\gamma;x)=N(J)\sum_{n=0}^{\infty}\frac{J^{n/2}e^{-i\E_n\gamma}}{\sqrt{\rho_n}}\psi_n^{(\varphi)}(x).
\label{33}\en
Here $J$ is a positive variable, $\gamma$ is real, and $\{\rho_n\}$ is a positive sequence with $\rho_0=1$ and $\rho_n=\E_n!=\E_1\E_2\cdots\E_n$, for $n\geq 1$. We refer to \cite{jpaeg} for some results on the un-rotated infinitely deep wall. In view of what we have already discussed before, and using in particular the fact that the eigenstates of $H_0$ and $H_\varphi$ coincide, most of the results deduced in \cite{jpaeg} also apply for us with minor modifications, as we will see.

In the definition (\ref{33}), $N(J)$ is a normalization which, in principle, could be not everywhere defined in $\mathbb{R}_+$, and could also depend on $\gamma$. However, this is not the case in the present situation. In fact, due to the orthonormalization condition $\langle\psi_k^{(\varphi)},\psi_n^{(\varphi)}\rangle_\varphi=\delta_{k,n}$, we find that $\|\Psi^{(\varphi)}\|_\varphi=1$ if and only if
$$
N(J)=\left(\sum_{n=0}^\infty\frac{J^n}{\rho_n}\right)^{-1/2},
$$
which is a power series everywhere convergent for $J\geq0$, as it can be easily checked using the explicit form for $\rho_n$. The series can be computed in terms of the Bessel function $I_2(x)$, \cite{grad}, and we find
$$
\sum_{n=0}^\infty\frac{J^n}{\rho_n}=\frac{8\pi^2}{JL^2}\,I_2\left(\frac{L\sqrt{J}}{\pi}\right).
$$
Now, let us introduce the function 
\be
\rho(J)=\left(\frac{L}{2\pi}\right)^4 JK_2\left(\frac{L\sqrt{J}}{\pi}\right),
\label{34}\en
where $K_2(x)$ is a modified Bessel function. In what follows we will need the following integral,
$$
\int_0^\infty t^{\mu-1}K_\nu(at)dt=2^{\mu-2}\,a^{-\mu}\Gamma\left(\frac{1}{2}(\mu+\nu)\right)\Gamma\left(\frac{1}{2}(\mu-\nu)\right),
$$
which holds if $\Re(\mu\pm\nu)>0$,
see \cite{magnus}. Of course, $\Gamma(x)$ is the Euler gamma function. With this choice of $\rho(J)$ in (\ref{34}), after some computation, we can check that $\rho(J)$ solves the following moment problem:
\be
\int_0^\infty J^n\rho(J)dJ=\rho_n.
\label{35}\en
Therefore, the resolution of the identity can be recovered: for all $f(x), g(x)\in\Hil_\varphi$ we have
\be
\int_C d\nu(J,\gamma)\langle f,\Psi^{(\varphi)}\rangle_\varphi \langle\Psi^{(\varphi)},g\rangle_\varphi= \langle f,g\rangle_\varphi,
\label{36}\en
where $C=\mathbb{R}_+\times\mathbb{R}$ and the measure $d\nu(J,\gamma)$ is defined on $C$ as follows:
$$
d\nu(J,\gamma)=N^{-2}(J)\rho(J)dJ\,d\nu(\gamma), \qquad\mbox{with}\qquad \int_{\mathbb{R}}\cdots d\nu(\gamma)=\lim_{\Gamma,\infty}\frac{1}{2\Gamma}\int_{-\Gamma}^{\Gamma}\cdots d\gamma,
$$
see \cite{gk}. It is easy to check also that the vectors $\Psi^{(\varphi)}(J,\gamma;x)$ in (\ref{33}) are temporarily stable: 
$$
e^{-ih_\varphi t}\Psi^{(\varphi)}(J,\gamma;x)=\Psi^{(\varphi)}(J,\gamma+t;x),
$$
and that
$$
\langle\Psi^{(\varphi)}(J,\gamma;x),h_\varphi\Psi^{(\varphi)}(J,\gamma;x)\rangle_\varphi=J.
$$
These are all properties useful, and natural, when dealing with coherent states. Moreover, see again \cite{gk}, it is possible to introduce a $(\gamma,\varphi)$-dependent lowering operator $a_\gamma^{(\varphi)}$ so that $\Psi^{(\varphi)}(J,\gamma;x)$ is an eigenstate of $a_\gamma^{(\varphi)}$. We put
$$
a_\gamma^{(\varphi)}\psi_n^{(\varphi)}(x)=\sqrt{\E_n}\,e^{i(\E_n-\E_{n-1})\gamma}\, \psi_{n-1}^{(\varphi)}(x),
$$
with the understanding that $a_\gamma^{(\varphi)}\psi_0^{(\varphi)}(x)=0$. Hence we get
$$
a_\gamma^{(\varphi)}\Psi^{(\varphi)}(J,\gamma;x)=\sqrt{J}\,\Psi^{(\varphi)}(J,\gamma;x).
$$

It can be checked  that the adjoint of $a_\gamma^{(\varphi)}$ in $\Hil_\varphi$, ${a_\gamma^{(\varphi)}}^{\dagger_\varphi}$, is a raising operator and, in general $[a_\gamma^{(\varphi)},{a_\gamma^{(\varphi)}}^{\dagger_\varphi}]\neq \1$: these operators do not satisfy the canonical commutation relations.

We end this section by noticing that, concerning coherent states, the situation is by far simpler than that for the Swanson model, where we were somehow forced to work with sets of functions which are complete in $\Lc^2(\mathbb{R})$, but which are not bases. Here we have used a different approach: all our sets are o.n. bases, but in different Hilbert spaces. This allows us to construct coherent states which appear of the same form as those in \cite{gk}, and there is no need to introduce what in the literature are called {\em bi-coherent states}, see \cite{bgs,bgs2} and references therein. Of course, there are pros and contras in both the approaches, but this depends on what in particular is interesting for us. For instance, introducing bi-coherent states as in \cite{bgs,bgs2}, we would get states which are not stable under time evolution, while those considered in this paper have this useful property. This is due to the fact that the eigenvalues of $h_\varphi$ do not depend linearly on the quantum number. However, bicoherent states satisfy all the main properties of ordinary coherent states, like the resolution of the identity or being eigenstates of some lowering operators. In particular, if the set $\F_{\Phi}^{(\varphi)}$ is a basis in $\Hil_0$, with biorthogonal basis $\F_{\tilde{\eta}}^{(\varphi)}$, then bicoherent states can be constructed in $\Hil_0$ either extending the Gazeau-Klauder procedure to non orthogonal bases, or adopting the general ideas proposed in  \cite{bgs,bgs2}.
In general, bicoherent states seem to be more relevant in connection with non self-adjoint Hamiltonians, which is not the situation we are dealing with here, as we have discussed in Section \ref{sectTH}.

\section{Conclusions}\label{sectconclu}

In this paper we have considered some aspects of the rotated IDSWP, introducing suitable o.n. bases in suitable Hilbert spaces, labeled by the rotation angle. We have seen how to define an unitary operator representing the rotation between different Hilbert spaces, and we have analyzed the properties of this operator in some details. 
We have also constructed the Gazeau-Klauder type coherent states and we have deduced some of their properties, and in particular we have proved the resolution of the identity, by solving the related moment problem.

We should stress that, despite what happens for the Swanson model, where the analysis is usually fully performed in $\Lc^2(\mathbb{R})$, here the use of $\Hil_\varphi$ seems more useful, and more user-friendly. 

It is not hard to imagine that the same idea, and the same rotation operator in particular, can be used to deform other quantum mechanical systems. This is an interesting approach, useful to create non self-adjoint Hamiltonians (in the original Hilbert space) with real eigenvalues, and with eigenvectors which can be bases or not, depending on our choice of the Hilbert space. The analysis of these systems is part of our future projects, together with a detailed analysis of the mathematical properties of Hilbert spaces like $\Lc^2\left(\Gamma_L\right)$, where $\Gamma_L=\{e^{i\varphi}x, x\in[-L/2,L/2]\}$, which could be seen as an alternative rotated version of $\Hil_0$, whose analysis seems deeply connected to analytic, square-integrable, functions. On the other hand, the same approach described in this paper seems to be easy to extend to deformations different from the rotations considered here or in the Swanson model, even if these are implemented by unbounded operators. In fact, this is what can be found, for instance, in many applications like those listed in \cite{baginbagbook}, or in \cite{bgv,ms}.

\section*{Acknowledgements}

This work was partially supported by the University of Palermo and by the Gruppo Nazionale di Fisica Matematica of Indam. The author thanks Prof. C. Trapani for several useful discussions during the preparation of this paper. The author also thanks Prof. J. Feinberg for interesting discussions during a preliminary preparation of this research.

\end{document}